\documentclass[preprint,5p,twocolumn,10pt]{elsarticle}

\usepackage[pdftex,hypertexnames=false,linktocpage=true]{hyperref}
\hypersetup{pdfauthor={Lin Bolang},colorlinks=true,linkcolor=red,anchorcolor=darkgreen,citecolor=green!50!blue,filecolor=darkgreen,urlcolor=green,bookmarksnumbered=true,pdfview=FitB}
\usepackage{lipsum}
\usepackage{mathtools,cuted}
\usepackage{cuted}

\usepackage{graphicx}
\usepackage{float}
\usepackage{subfigure}

\usepackage{array}
\usepackage{booktabs} 
\usepackage{arydshln}
\usepackage{multirow}
\usepackage{makecell} 

\newcommand{\be}{\begin{equation}}  
\newcommand{\ee}{\end{equation}}  

\newcommand{\ba}{\begin{align}}  
\newcommand{\ea}{\end{align}}

\newcommand{\ra}{\rangle}
\newcommand{\la}{\langle}

\newcommand{\nn}{\nonumber}

\hypersetup{colorlinks=true, citecolor=blue, urlcolor=blue, linkcolor=blue}
\newcommand*{\dif}{\mathop{}\!\mathrm{d}} 

\usepackage{braket}

\usepackage{comment}
\begin{document}

\begin{frontmatter}
\title{Chiral-odd gluon generalized parton distributions in the proton: A light-front quantization approach}


\author[imp,ucas,keylab]{Bolang~Lin}
\ead{linbolang@impcas.ac.cn}

\author[imp,ucas,keylab]{Sreeraj~Nair\corref{cor1}}
\ead{sreeraj@impcas.ac.cn}

\author[imp,ucas,keylab]{Chandan~Mondal}
\ead{mondal@impcas.ac.cn}

\author[imp,ucas,keylab]{Siqi~Xu}
\ead{xsq234@impcas.ac.cn}

\author[hearo]{Zhi~Hu}
\ead{huzhi0826@gmail.com}

\author[lzu]{Pengxiang~Zhang}
\ead{zhangpx2021@lzu.edu.cn}

\author[imp,ucas,keylab]{Xingbo~Zhao}
\ead{xbzhao@impcas.ac.cn}

\author[iowa]{James~P.~Vary}
\ead{jvary@iastate.edu}

\author[]{\\\vspace{0.2cm}(BLFQ Collaboration)}

\address[imp]{Institute of Modern Physics, Chinese Academy of Sciences, Lanzhou, Gansu, 730000, China}
\address[ucas]{School of Nuclear Physics, University of Chinese Academy of Sciences, Beijing, 100049, China}
\address[keylab]{CAS Key Laboratory of High Precision Nuclear Spectroscopy, Institute of Modern Physics, Chinese Academy of Sciences, Lanzhou 730000, China}
\address[hearo]{High Energy Accelerator Research Organization (KEK), Ibaraki 305-0801, Japan}
\address[lzu]{Lanzhou University, Lanzhou, Gansu, 730000, China}
\address[iowa]{Department of Physics and Astronomy, Iowa State University, Ames, IA 50011, USA}
\cortext[cor1]{Corresponding author}

\begin{abstract}
Within the basis light-front quantization (BLFQ) framework, we evaluate the gluon chiral-odd generalized parton distributions (GPDs) inside the proton at zero skewness. We employ the light-front wave functions of the proton obtained from a light-front quantized Hamiltonian with quantum chromodynamics input using BLFQ. Our investigation encompasses both the valence Fock sector with three constituent quarks and an additional sector containing three quarks and a dynamical gluon. We analyze the gluon GPDs in the momentum space as well as in the transverse position space. We further present the gluon's generalized form factors derived from the Mellin moments of its chiral-odd GPDs. Using the proton transverse spin sum rule, we also present the $x$-dependence of the angular momentum carried by the polarized gluon and determine the relative contributions of quarks and the gluon to the transversity asymmetry.
\end{abstract}
\end{frontmatter}

\section{Introduction\label{Sec1}}
Illuminating the three-dimensional structure of nucleons through the lens of their constituent quarks and gluons stands as a formidable challenge in modern particle and nuclear
physics. Parton distribution functions (PDFs) are fundamental in exploring the structure of hadrons, encoding
the distribution of longitudinal momentum and polarization
carried by the constituents.
While PDFs are extensively employed, achieving a detailed depiction of hadrons' nonperturbative structure necessitates expanding these functions into more comprehensive, higher-dimensional distributions, known as generalized parton distributions (GPDs)~\cite{Belitsky:2005qn,Diehl:2003ny,Ji:1996nm,Radyushkin:1997ki,Muller:1994ses}. 
GPDs depend on three essential parameters: the parton's share of the hadron's longitudinal momentum ($x$), the skewness ($\zeta$), which quantifies the longitudinal momentum fraction exchanged in the interaction, and the square of the total momentum exchange ($t$).
GPDs serve to merge the insights provided by PDFs with the spatial insights offered by form factors (FFs), thereby presenting a cohesive distribution that encapsulates both the momentum and spatial dimensions of hadronic structure~\cite{Guidal:2004nd,SattaryNikkhoo:2018gzm,Miller:2007uy}. 
When the skewness $\zeta$ is zero, the Fourier transform of the GPDs with respect to the transverse momentum transfer ($\Delta_\perp$) results in impact parameter-dependent parton distribution functions (ipdpdfs)~\cite{Burkardt:2000za,Burkardt:2002hr}. These distributions map how partons with given longitudinal momenta are distributed in the transverse position or impact parameter space, $b_\perp$. In contrast to GPDs, ipdpdfs have a probabilistic interpretation and follow positivity constraints, offering insight into the partons' spatial configuration within a hadron~\cite{Ralston:2001xs}.

Unlike PDFs, which are extracted from inclusive processes like deep-inelastic scattering between leptons and hadrons, GPDs are explored via exclusive methods, primarily through deeply virtual Compton scattering (DVCS) and deeply virtual meson production (DVMP)~\cite{Ji:1996nm,Radyushkin:1997ki,Diehl:2015uka,Radyushkin:1996nd,Radyushkin:1996ru,Collins:1996fb}, timelike Compton scattering (TCS)~\cite{Berger:2001xd}, single diffractive hard exclusive processes (SDHEPs)~\cite{Qiu:2022pla,Grocholski:2022rqj,Duplancic:2022ffo}, wide-angle Compton scattering (WACS)~\cite{Radyushkin:1998rt,Diehl:1998kh}, and also double DVCS~\cite{Deja:2023ahc}. The exclusive photoproduction of a photon-meson pair provides access to both chiral-even and chiral-odd GPDs, depending on the polarization of the outgoing meson~\cite{Duplancic:2023kwe}. Accessing GPDs is a key focus for the upcoming Electron-Ion Collider (EIC)~\cite{Accardi:2012qut,AbdulKhalek:2021gbh} at Brookhaven National Laboratory (BNL) and the upgrade at Jefferson Lab (JLab)~\cite{Accardi:2023chb}. JLab has already made significant contributions through exclusive measurements from extensive datasets~\cite{CLAS:2018ddh,CLAS:2021gwi,Georges:2017xjy,JeffersonLabHallA:2022pnx}. With the EIC and similar initiatives like the Electron-Ion Collider in China (EicC)~\cite{Anderle:2021wcy}, the field is poised for a significant increase in data revealing details of the proton's GPDs.

From the theoretical perspective, a range of nonperturbative techniques has been employed to delve into quark GPDs, adopting a more phenomenological approach to their study~\cite{Pasquini:2005dk,Pasquini:2006dv,Meissner:2009ww,Boffi:2002yy,Scopetta:2003et,Choi:2001fc,Choi:2002ic}. Meanwhile, promising theoretical frameworks for obtaining GPDs also include the  Euclidean lattice  QCD~\cite{Ji:2013dva,Ji:2020ect,Lin:2021brq,Lin:2020rxa,Bhattacharya:2022aob, Alexandrou:2021bbo, Alexandrou:2022dtc, Guo:2022upw, Alexandrou:2020zbe, Gockeler:2005cj, QCDSF:2006tkx, Alexandrou:2019ali}. Investigation into gluon GPDs has not been as extensive as that into quark GPDs. Recent studies have increasingly addressed both chiral even and chiral odd gluon GPDs, with a particular interest in using light-cone spectator models to examine these leading twist phenomena~\cite{Tan:2023kbl,Chakrabarti:2024hwx}. 

At leading twist, the proton features eight gluon GPDs. Specifically, four are chiral even: $H_g$, $E_g$, $\tilde{H}_g$, and $\tilde{E}_g$; and the other four are chiral odd: $H_{T}^g$, $E_{T}^g$, $\tilde{H}_{T}^g$, and $\tilde{E}_{T}^g$, distinguished by their chiral properties~\cite{Diehl:2003ny}. In our preceding study, we explored the chiral even GPDs~\cite{Lin:2023ezw}. Building on that foundation, the current study focuses on deriving the chiral-odd gluon GPDs at zero skewness, employing the basis light-front quantization (BLFQ) methodology. The BLFQ framework, a nonperturbative approach, effectively addresses relativistic many-body problems in quantum field theories~\cite{Vary:2009gt,Zhao:2014xaa,Nair:2022evk,Wiecki:2014ola,Li:2015zda,Jia:2018ary,Tang:2019gvn,Lan:2019vui,Mondal:2019jdg,Xu:2021wwj,Lan:2021wok,Kuang:2022vdy}. Within BLFQ, an effective light-front Hamiltonian is employed to determine mass eigenstates, incorporating the nucleon Fock sector with one dynamical gluon  ($\ket{qqqg}$) beyond the three-quark valence Fock sector ($\ket{qqq}$)~\cite{Xu:2023nqv,Yu:2024mxo,Zhu:2024awq}. This framework incorporates the QCD interaction applicable to both sectors.We explore the chiral-odd gluon GPDs within the impact parameter space, making a transition from transverse momentum to impact parameter space through Fourier transformation. Additionally, the analysis includes assessing the contributions of quarks and gluons to the proton's transversity asymmetry by examining moments of GPDs.

\section{BLFQ methodology: Modeling the proton with a single dynamical gluon\label{Sec2}}
In the realm of light-front\footnote{We adopt the light-front convention for the four-vector $a=\left( a^+, a^- , a^{\perp}\right)$, where $a^{\pm} = a^0 \pm a^3$ and $\vec{a}_{\perp} = \left(a^1,a^2 \right)$.} quantum field theory, the determination of bound states is anchored in the solution of the mass eigenvalue equation:
\begin{equation}
	(P^+ P^- - \vec{P}_\perp^{~2})\ket{\Psi} = M^2 \ket{\Psi}
\end{equation}
with $P^-$ being the light-front Hamiltonian, $P^+$ the longitudinal momentum, $\vec{P}_\perp$ the transverse momentum, and $M$ the invariant mass of the system. 
We proceed to express the proton's bound state in Fock space, limiting our expansion to the Fock sector that includes a single dynamical gluon, outlined schematically  as,
\begin{equation}
	\ket{\Psi}=\psi_{qqq}\ket{qqq}+\psi_{qqqg}\ket{qqqg}+\cdots,
	\label{fockeq}
\end{equation}
where $\psi_{\cdots}$ signifies the light-front wave functions (LFWFs) linked to each Fock state $\ket{\cdots}$. For practical computational purposes, it becomes essential to limit the expansion of the Fock sectors to a manageable number of dimensions.
At the model scale, we represent the proton as the light-front wave functions for the valence quarks, denoted as $\psi_{uud}$, along with configurations that include three quarks and an additional dynamical gluon, symbolized as $\psi_{uudg}$.

Building on the Fock sector truncation retaining the first two terms in Eq.~(\ref{fockeq}), our approach explicitly includes both quarks and gluons as fundamental degrees of freedom and utilizes an effective light-front Hamiltonian, $P^- = P^-_0 + P^-_I$. Here, $P^-_0$ refers to the light-front QCD Hamiltonian, specifically applied to the $\ket{qqq}$ and $\ket{qqqg}$ Fock states. $P^-_I$, on the other hand, represents a model Hamiltonian dedicated to capturing the confining dynamics~\cite{Xu:2022abw}. In the light-front gauge $A^+ = 0$, the light-front QCD
Hamiltonian with one dynamical gluon is given by~\cite{Xu:2023nqv,Lan:2021wok},
\begin{equation}
	\begin{split}
		P_0^-= &\int \mathrm{d}x^- \mathrm{d}^2 x^{\perp} \Big\{\frac{1}{2}\bar{\psi}\gamma^+\frac{m_{0}^2+(i\partial^\perp)^2}{i\partial^+}\psi \\
		&+\frac{1}{2}A_a^i\left[m_g^2+(i\partial^\perp)^2\right] A^i_a +g_c\bar{\psi}\gamma_{\mu}T^aA_a^{\mu}\psi \\
		&+ \frac{1}{2}g_c^2\bar{\psi}\gamma^+T^a\psi\frac{1}{(i\partial^+)^2}\bar{\psi}\gamma^+T^a\psi \Big\}.
	\end{split}
	\label{hamieq}
\end{equation}
The initial two terms in Eq.~(\ref{hamieq}) describe the kinetic energy of quark and gluon, characterized by their respective bare masses, $m_0$ for quark and $m_g$ for gluon. Here, $\psi$ signifies the quark field, and $A^i_a$ represents the gluon field. The Dirac matrices are denoted by $\gamma^\mu$, and $T$ stands for the SU$(3)$ gauge group's generator in color space. Despite gluons being massless in standard QCD, our model introduces a phenomenological gluon mass to better account for low-energy phenomena as highlighted in Ref.~\cite{Xu:2022abw}. The third and fourth terms in Eq.~(\ref{hamieq}) detail the interactions between quark and gluon fields through a vertex and instantaneous gluon interactions, both controlled by the coupling constant $g_c$. In line with Fock-sector dependent renormalization approaches~\cite{Perry:1990mz,Karmanov:2008br,Kurganov:2024tdn}, a mass counterterm ($\delta m_q$) is implemented to adjust the quark mass within the leading Fock sector to a renormalized value, $m_q = m_0 - \delta m_q$. Additionally, the model allows for an independent quark mass $m_f$ in the vertex interactions following Refs.~\cite{Glazek:1992aq,Burkardt:1998dd}.

In our framework, the implementation of confinement for the $\ket{qqq}$ Fock state is given by the following equation:
\begin{align}
	P^-_IP^+=\frac{\kappa^4}{2}\sum\limits_{i\neq j}\left\{\Vec{r}_{ij\perp}^{\,2}-\frac{\partial_{x_i}(x_ix_j\partial_{x_j})}{(m_i+m_j)^2}\right\},
\end{align}
in which $\Vec{r}_{ij\perp}$ serves as the holographic variable~\cite{Brodsky:2014yha} that quantifies the transverse distance between partons, and $\kappa$ represents the strength of the confinement. For the $\ket{qqqg}$ Fock state, our model omits an explicit confinement term, relying instead on the effective mass of the gluon and the chosen basis to account for the effects of confinement.

For the longitudinal basis, we use plane waves confined within a one-dimensional box of length $2L$, with antiperiodic and periodic boundary conditions for fermions and bosons, respectively. Specifically, the longitudinal momentum is discretized as $p_i^+=2\pi k_i/L$, where $k_i$ represents the longitudinal quantum number. $k_i$ is a half-integer
(integer) for fermions (bosons). Here, we omit the zero
mode of bosons. In the transverse plane, we utilize two-dimensional harmonic oscillator (2D-HO) basis functions, $\Phi_{nm}(\Vec{p}_\perp;b)$, with $n$ and $m$ indicating the principal and orbital momentum quantum numbers, respectively, and $b$ setting the energy scale~\cite{Li:2015zda}. The light-front helicity is denoted by $\lambda$. Each particle in our Fock-state basis is described by four quantum numbers $|\alpha_i\rangle=|k_i,n_i,m_i,\lambda_i\rangle$, and multi-particle states are constructed as their direct product $|\alpha\rangle=\otimes_i |\alpha_i\rangle$, ensuring a total color singlet configuration.
All the Fock-sector basis states have the same total angular momentum projection $M_j$ such that $\sum_i(\lambda_i+m_i)=M_j$.

We introduce cutoffs, $\mathcal{K}$ and $\mathcal{N}_{\mathrm{max}}$, to limit the basis in the longitudinal and transverse directions, respectively. The total longitudinal momentum of the system is represented by the dimensionless variable $\mathcal{K}=\sum_{i}k_i$, from which we derive the longitudinal momentum fraction of each parton as $x_i=k_i/\mathcal{K}$. The transverse energy cutoff, $\mathcal{N}_{\mathrm{max}}$, defined as $\mathcal{N}_{\mathrm{max}}\geq\sum_i(2n_i+|m_i|+1)$, dictates the range of the 2D-HO states included and implies infrared (IR) and ultraviolet (UV) cutoffs, approximated as $\lambda_{\mathrm{IR}}\approx b/\sqrt{\mathcal{N}_{\mathrm{max}}}$ and $\lambda_{\mathrm{UV}}\approx b\sqrt{\mathcal{N}_{\mathrm{max}}}$, respectively~\cite{Zhao:2014xaa}.

Diagonalizing the light-front Hamiltonian matrix within this framework yields the mass-squared spectra, $M^2$, and the LFWFs in momentum space:
\begin{equation}\label{wfs}
	\Psi^{M_J}_{N,\{\lambda_i\}}(\{x_i,\vec{p}_{i\perp}\})=\sum_{\{n_i m_i\}}\psi^{M_J}_{N}(\{{\alpha}_i\})\prod_{i=1}^{N}  \Phi_{n_i m_i}(\vec{p}_{i\perp},b),
\end{equation}
linking the $\psi^{M_J}_{N=3}({\alpha_i})$ and $\psi^{M_J}_{N=4}({\alpha_i})$ components to the $\ket{uud}$ and $\ket{uudg}$ sectors, respectively.

\begin{figure*}[hbt!]
	\begin{center}
		\includegraphics[width=0.32\linewidth]{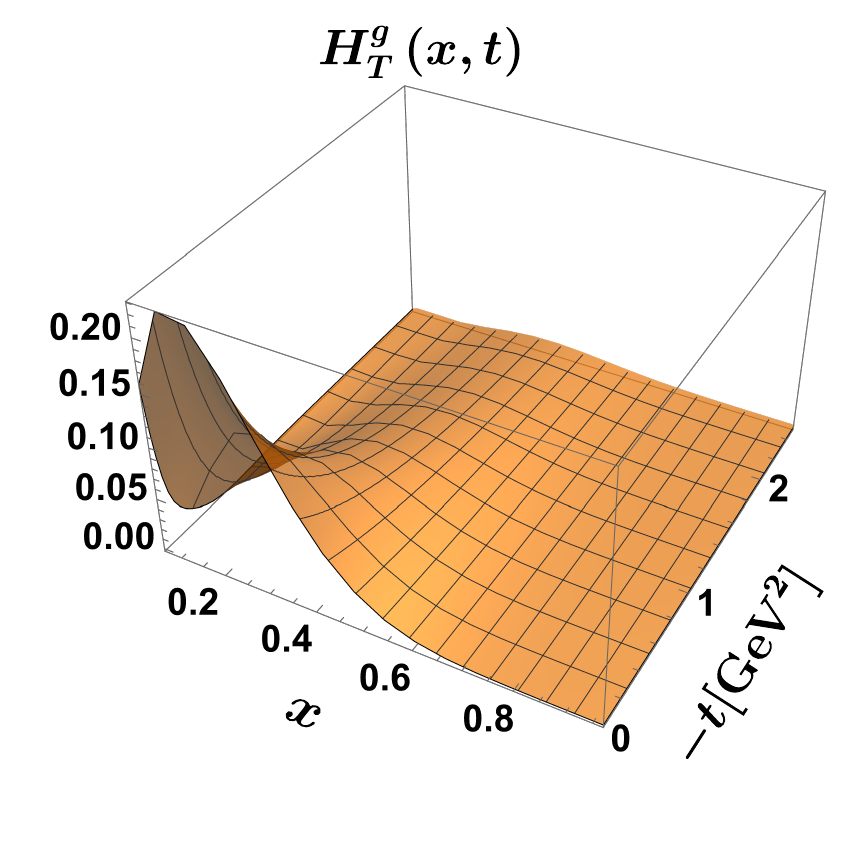}
		\includegraphics[width=0.32\linewidth]{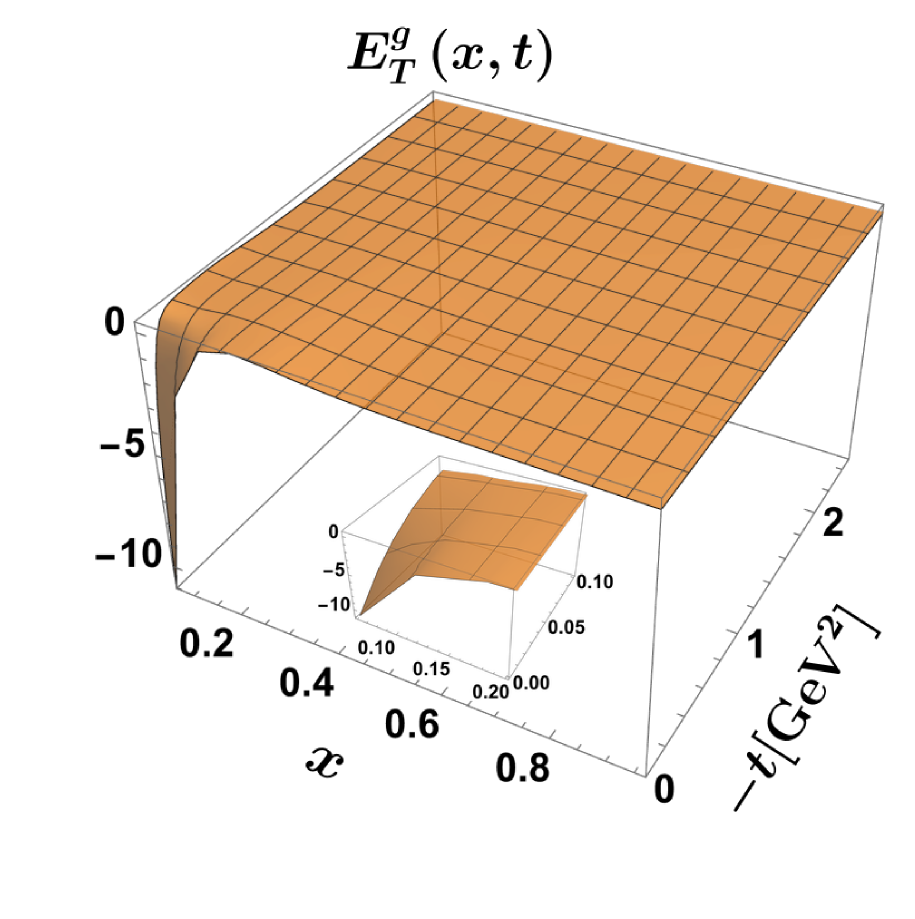}
		\includegraphics[width=0.32\linewidth]{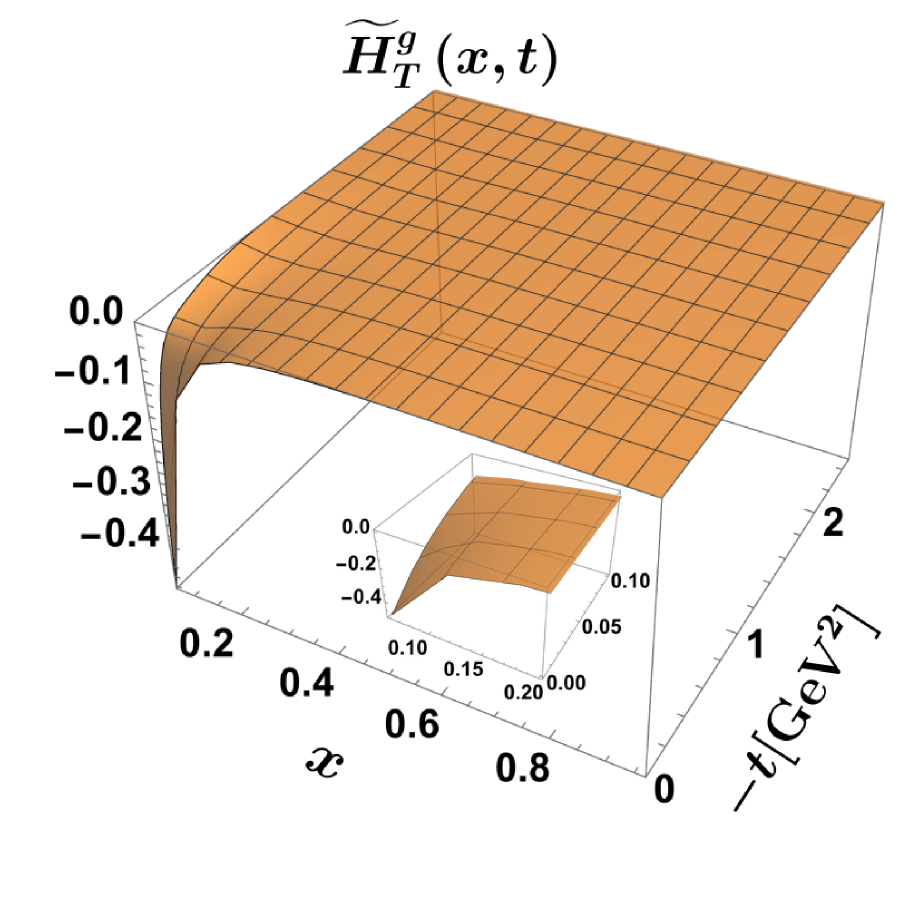}
		\caption{The chiral-odd gluon GPDs at zero skewness, left: $H_T^g(x,t)$, middle: $E_T^g(x,t)$, and right: $\widetilde{H}_T^g(x,t)$ as functions of $x$ and $-t$. The GPDs are evaluated at $\mathcal{N}_{\mathrm{max}}=9$, $\mathcal{K}=16.5$.}
		\label{gluon GPD in momentum space}
	\end{center}
\end{figure*}
\section{Gluon generalized parton distributions\label{Sec3}}
At leading twist, there are four  chiral-odd (helicity flip) gluon GPDs denoted by $H_T^g$, $E_T^g$, $\widetilde{H}_T^g$ and $\widetilde{E}_T^g$. Unlike ordinary parton distributions, where gluon helicity flip requires spin-$1$ or higher targets for angular momentum conservation,  GPDs do not have this limitation. They facilitate transverse momentum transfer and orbital angular momentum changes. These GPDs are defined by the matrix element of the bilocal gluon tensor operator $\mathbf{S}F^{+i}(-z/2)F^{+j}(z/2)$, where $\mathbf{S}$ indicates symmetrization and trace subtraction~\cite{Diehl:2001pm}. Here $F^{+\mu}(x) = \partial^{+}A^{\mu}(x)$ is the gluon field tensor in the light cone gauge $A^+ = 0$. Using a compact notation $[a,b]^{+i} = a^+ b^i - b^+ a^i$, the chiral-odd gluon GPDs can be parametrized as follows:
\begin{align}
	\label{gpdoddeq}
 -\frac{1}{{P}^+}&\int \frac{{\rm d}z^-}{2\pi} e^{ixP^+z^-} 
	\la P',\Lambda' | \mathbf{S} F^{+i} \big(-\frac{z}{2}\big) F^{+j} \big(\frac{z}{2} \big) | P,\Lambda \ra \Big{|}_{\substack{z^+=0\\z^{\perp}=0}} \nn \\
	&= \frac{\mathbf{S}}{2 {P}^{+}} \frac{[{P},\Delta]^{+j}}{2 M{P}^+}\bar{u}( P^{\prime}, \Lambda^{\prime}) \Bigg[i \sigma^{+i} H_T^g+\frac{[\gamma, \Delta]^{+i} }{2 M} E_T^g \nn \\  
	&+ \frac{[{P}, \Delta]^{+i} }{M^2} \widetilde{H}_T^g +\frac{[\gamma,{P}]^{+i} }{M} \widetilde{E}_T^g \Bigg] u(P, \Lambda).
\end{align}
 We omit the arguments $(x,\zeta,t)$ of the GPDs in Eq.~\eqref{gpdoddeq}.
 We denote the initial and final proton states by momenta $P$ and $P'$, and helicities $\Lambda$ and $\Lambda'$, respectively. $u(\bar{u})$ represents the spinor. 
We choose a frame such that the momenta of the target proton at the initial and final state, at $\zeta=0$, become 
\begin{align}
P &= \left(P^+,\vec{0}_\perp, \frac{M^2}{P^+}  \right),\\
P^{\prime} &= \left(P^+ ,- \vec{\Delta}_\perp,\frac{M^2+\vec{\Delta}_\perp^2}{P^+} \right).
\end{align} 
Our focus narrows to the case of zero skewness, targeting the DGLAP region $\zeta < x < 1$.  Time reversal invariance shows that $\widetilde{E}_T^g$ changes sign with $\zeta$, i.e., $\widetilde{E}_T^g(x,\zeta,t) = - \widetilde{E}_T^g(x,-\zeta,t)$, making non-zero skewness essential for calculating $\widetilde{E}_T^g$. Using the reference frame where the momenta $\vec{P}_{\perp}$ and $\vec{P}^\prime_{\perp}$ lie in the $x$-$z$ plane, the remaining three non-zero chiral-odd gluon GPDs at zero skewness are depicted through the diagonal overlap of LFWFs as follows:
\begin{equation}\label{gpd_g_eq1}
\begin{split}
&H^g_T\left( x,0,t\right) =-\frac{2M}{{\Delta}_\perp} \sum_{ \{\lambda_i\}} \int\left[\dif\mathcal{X}\dif\mathcal{P}_\perp\right] \delta(x-x_1) \\
&\times \left[ \Psi_{ 4,\{\lambda_1,\lambda_{2\cdots4} \}}^{\downarrow*} (\{x^\prime_i,\vec{p}^{\,\prime}_{i\perp}\}) \Psi_{ 4,\{-\lambda_1,\lambda_{2\cdots4} \}}^{\uparrow} (\{x_i,\vec{p}_{i\perp}\}) \right. \\
&+ \left. \Psi_{ 4,\{\lambda_1,\lambda_{2\cdots4} \}}^{\uparrow*} (\{x^\prime_i,\vec{p}^{\,\prime}_{i\perp}\}) \Psi_{ 4,\{-\lambda_1,\lambda_{2\cdots4} \}}^{\downarrow} (\{x_i,\vec{p}_{i\perp}\})\right],
\end{split}
\end{equation}
\begin{equation}\label{gpd_g_eq2}
\begin{split}
&E^g_T\left(x,0,t\right) = \frac{4M^2}{\Delta_\perp^2}\sum_{ \{\lambda_i \}} \int\left[\dif\mathcal{X}\dif\mathcal{P}_\perp\right] \delta(x-x_1) \\
&\times \Big[ \Psi_{ 4,\{\lambda_1,\lambda_{2\cdots4} \}}^{\downarrow*} (\{x^\prime_i,\vec{p}^{\,\prime}_{i\perp}\}) \Psi_{ 4,\{-\lambda_1,\lambda_{2\cdots4} \}}^{\downarrow} (\{x_i,\vec{p}_{i\perp}\}) \\
&+ \Psi_{ 4,\{\lambda_1,\lambda_{2\cdots4} \}}^{\uparrow*} (\{x^\prime_i,\vec{p}^{\,\prime}_{i\perp}\}) \Psi_{ 4,\{-\lambda_1,\lambda_{2\cdots4} \}}^{\uparrow} (\{x_i,\vec{p}_{i\perp}\})  \\
&- \frac{4M}{\Delta_\perp}\Psi_{ 4,\{\lambda_1,\lambda_{2\cdots4} \}}^{\downarrow*} (\{x^\prime_i,\vec{p}^{\,\prime}_{i\perp}\}) \Psi_{ 4,\{-\lambda_1,\lambda_{2\cdots4} \}}^{\uparrow} (\{x_i,\vec{p}_{i\perp}\})\Big],
\end{split}
\end{equation}
\begin{equation}\label{gpd_g_eq3}
\begin{split}
&\widetilde{H}^g_T\left(x,0,t\right) = \frac{8M^3}{\Delta_\perp^3} \sum_{ \{\lambda_i\}} \int\left[\dif\mathcal{X}\dif\mathcal{P}_\perp\right] \delta(x-x_1) \\
&\times \Psi_{ 4,\{\lambda_1,\lambda_{2\cdots4} \}}^{\downarrow*} (\{x^\prime_i,\vec{p}^{\,\prime}_{i\perp}\}) \Psi_{ 4,\{-\lambda_1,\lambda_{2\cdots4} \}}^{\uparrow} (\{x_i,\vec{p}_{i\perp}\}) ,
\end{split}
\end{equation}
where the light-front momenta of the final struck parton are \(x'_1=x_1\) and  $\vec{p}^{\,\prime}_{1\perp} ={p}_{1\perp} - (1-x_1)\vec{\Delta}_{\perp}$. For the spectators these momenta become $x_i^\prime = x_i$ and $\vec{p}_{i\perp}^{\,\prime} = \vec{p}_{i\perp} + x_i \vec{\Delta}_\perp(i=2,3,4)$. Here, $t=-\vec{\Delta}_\perp^2$, when the skewness $\zeta=0$.
The integration measure appearing in Eqs.~\eqref{gpd_g_eq1}-\eqref{gpd_g_eq3} is succinctly represented by the following shorthand notation:
\begin{equation}
	\begin{split}
		\left[\dif\mathcal{X}\dif\mathcal{P}_\perp\right]&\equiv \prod_{i=1}^N\left[\frac{\dif x_i \dif^2\vec{p}_{i\perp}}{16\pi^3}\right]16\pi^3\delta\left(1-\sum_{i=1}^N x_i\right)\\
		&\times \delta^2\left(\sum_{i=1}^N\vec{p}_{i\perp}\right).
	\end{split}
\end{equation}

The study of GPDs is enhanced by analyzing their moments, particularly $x$-moments.
By evaluating these moments at zero momentum transfer, it becomes possible to quantify the contribution of quarks and gluons to the hadron's momentum, spin, and angular momentum. The  first two Mellin moments corresponding to $n=1$ and $n=2$ of the chiral-odd gluon GPDs at zero skewness can be defined as~\cite{Kaur:2023lun},
\begin{equation}
	\begin{split}
 A^g_{Tn0}&=\int {\rm d}x\,x^{n-1} H^g_T(x,0,t), \\
 B^g_{Tn0}&=\int {\rm d}x\, x^{n-1}  \left(E^g_T(x,0,t)+2\widetilde{H}^g_T(x,0,t)\right),
 \label{mellin}
	\end{split}
\end{equation}
where the zero added to the subscript specifies that the moments are calculated at zero skewness. The terms $A^g_{Tn0}$ and $B^g_{Tn0}$ represent the generalized form factors.
 Notably, the sum $E^g_T +2\widetilde{H}^g_T$ is considered to offer more fundamental insights than $E^g_T$ alone~\cite{Diehl:2005jf}. This specific combination, $E^g_T +2\widetilde{H}^g_T$, is linked to how the transverse spin and angular momentum are correlated within an unpolarized proton. 
The second moments of chiral-odd GPDs are related with the angular momentum carried by the constituents with the transverse spin along the $\widehat{x}$ direction in an unpolarized proton. This quantity is one-half of the expectation value of the transversity asymmetry defined as~\cite{Burkardt:2005hp},
\begin{align}
	\braket{\delta^i J^i}&=\frac{1}{2} \int {\rm d}x \,x \left[H_T(x,0,0)+E_T(x,0,0)+2\widetilde{H}_T(x,0,0)\right] \nn \\
	&=\frac{1}{2}\left[A_{T20}(0)+B_{T20}(0)\right],\label{gluon ta}
\end{align}
which is the analogue of Ji sum rule~\cite{Ji:1996ek},
\begin{align}
J^z &= \frac{1}{2} \int {\rm d}x \,x \left[H(x,0,0) +E(x,0,0)\right]\nn \\ &= \frac{1}{2}\left[A_{20}(0)+B_{20}(0)\right] \;.
\label{eq:angular-momentum}
\end{align}

Our analysis extends to examining gluon GPDs in the impact parameter space, accomplished by conducting a two-dimensional Fourier transform (FT) relative to the transverse momentum transfer, specifically at zero skewness. The 
GPDs in impact parameter space are defined as,
\begin{equation}
  \mathcal{F}\left(x,b_{\perp}\right)=\int \frac{\dif^2\vec{\Delta}_\perp}{\left(2\pi\right)^2}e^{-i \vec{\Delta}_\perp \cdot \vec{b}_{\perp}}F\left(x,t\right),
\end{equation}
where $\mathcal{F}(x,b_{\perp})$ denotes the GPD mapped into  impact parameter space. Here, $\vec{b}_{\perp}$ represents the impact parameter, the Fourier conjugate of the transverse momentum transfer $\vec{\Delta}_{\perp}$ within the plane perpendicular to the particle's motion. The variable $b_{\perp}=|\vec{b}_{\perp}|$ also describes the transverse distance between the active parton and the proton's center of momentum, satisfying the condition $\sum_i x_i \vec{b}_{\perp i} = 0$, summed over all partons~\cite{Diehl:2003ny}.

 \begin{figure}[!ht]
	\begin{center}
		\includegraphics[width=0.95\linewidth]{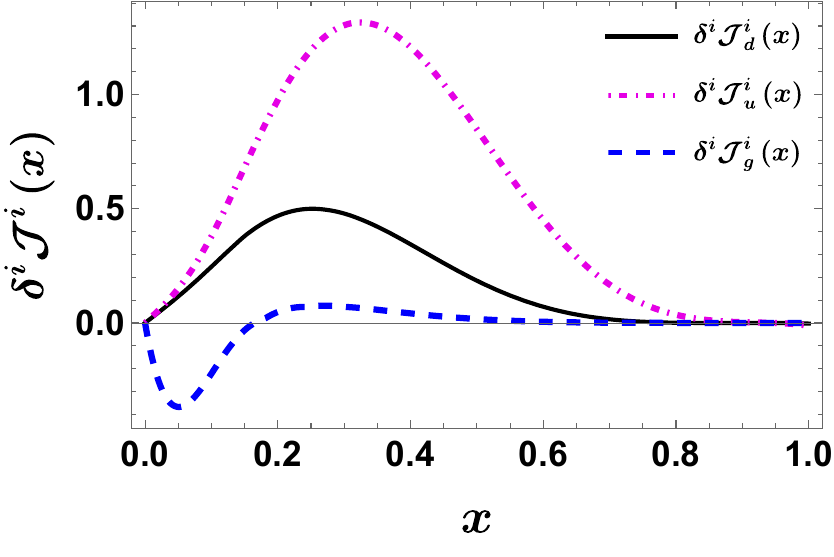}
		\caption{The $x$-unintegrated transversity asymmetry: $\delta^i \mathcal{J}^i(x)=x [H_T(x,0)+E_T(x,0)+2\widetilde{H}_T(x,0)]$ as a function of $x$. The $u$-quark and $d$-quark contributions are shown by the dot-dashed magenta and solid black lines, respectively. Gluon contribution is shown by the dotted blue line. The results are evaluated at $\mathcal{N}_{\mathrm{max}}=9$, $\mathcal{K}=16.5$.}
	\label{transverse asymmetry}
	\end{center}
\end{figure}
\begin{figure}[!ht]
	\begin{center}
		\includegraphics[width=0.95\linewidth]{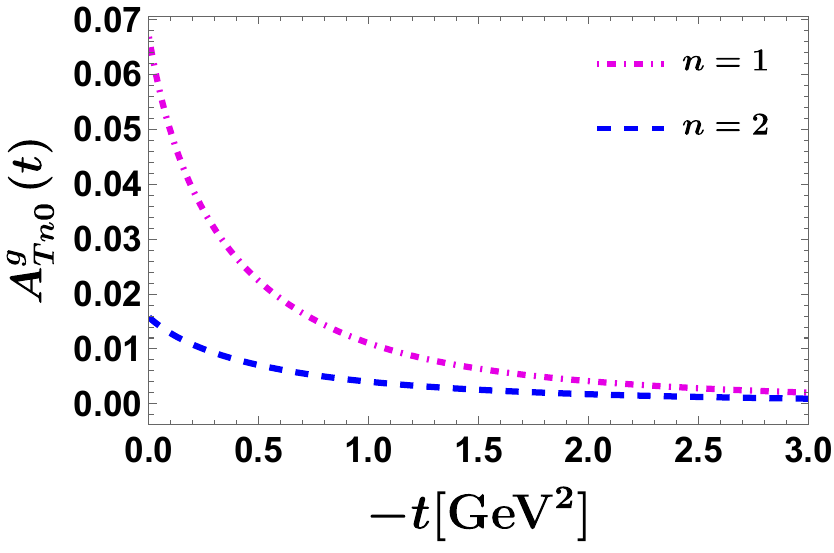}
		\includegraphics[width=0.95\linewidth]{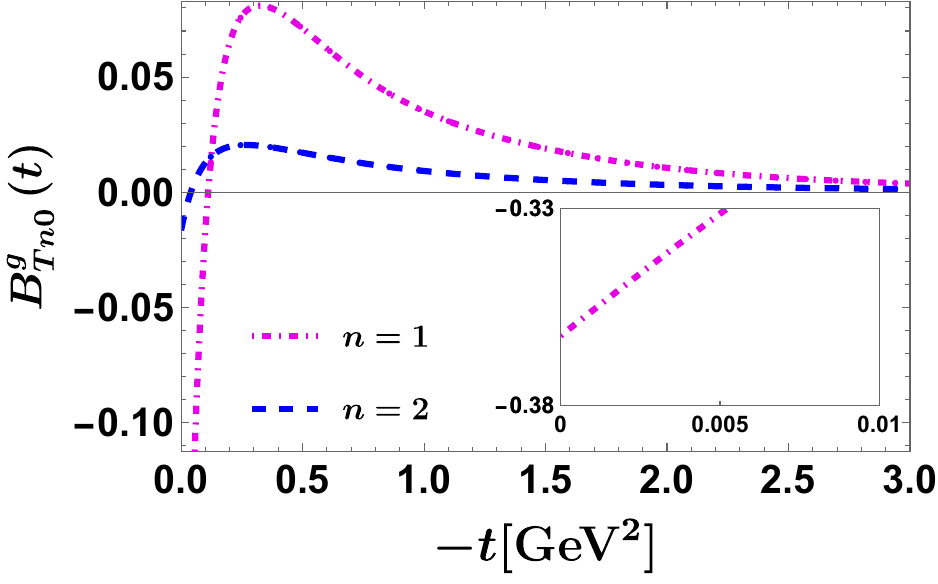}
		\caption{The first moment (dot-dashed magenta) and the second moment (dashed blue) of the chiral-odd GPDs, $H_T^g(x,t)$ (upper panel) and $E_T^g(x,t)+2\widetilde{H}_T^g(x,t)$ (lower panel)  as functions of $-t$ (in units of GeV$^2$). The results are evaluated at $\mathcal{N}_{\mathrm{max}}=9$, $\mathcal{K}=16.5$.}
\label{Fig:moments}
	\end{center}
\end{figure}
\begin{figure*}[htp]
	\begin{center}
		\includegraphics[width=0.32\linewidth]{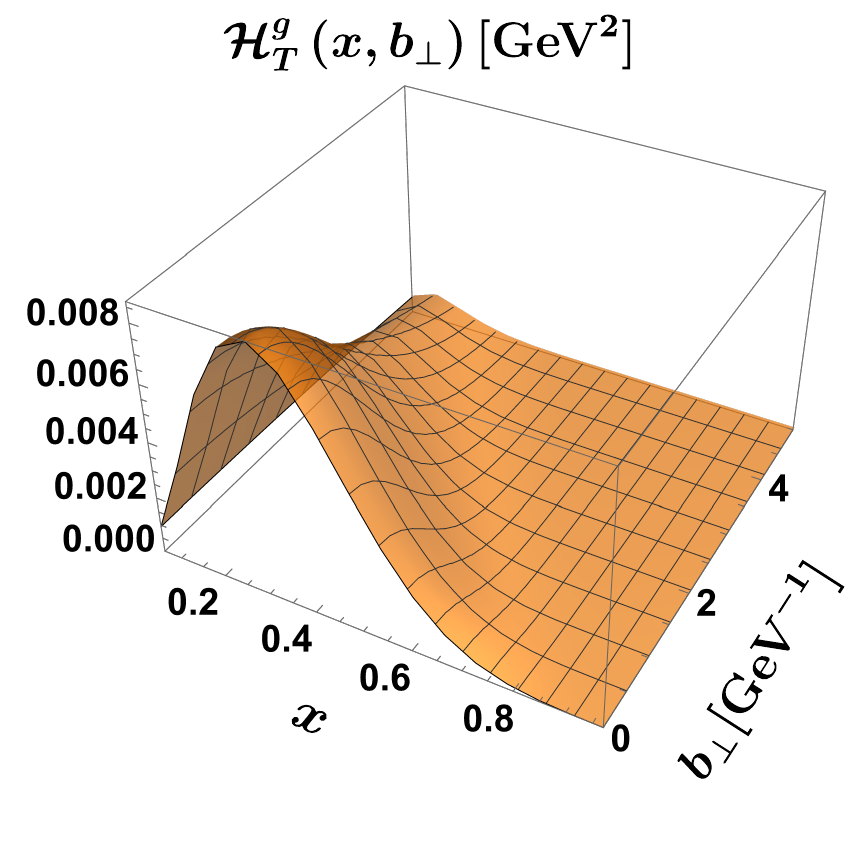}
		\includegraphics[width=0.32\linewidth]{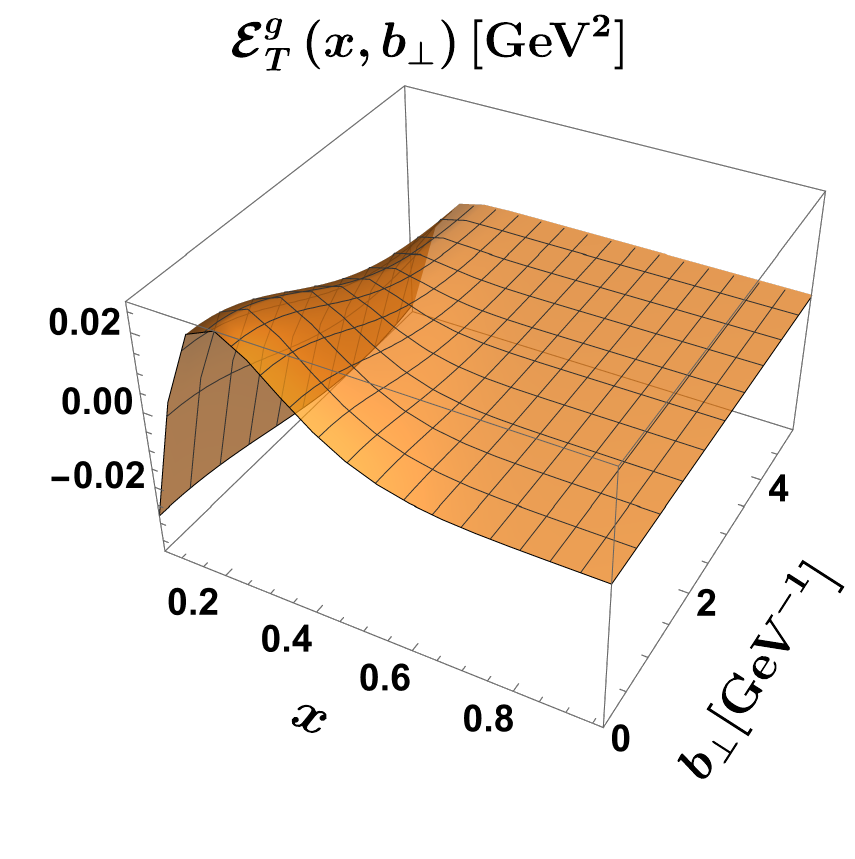}
		\includegraphics[width=0.32\linewidth]{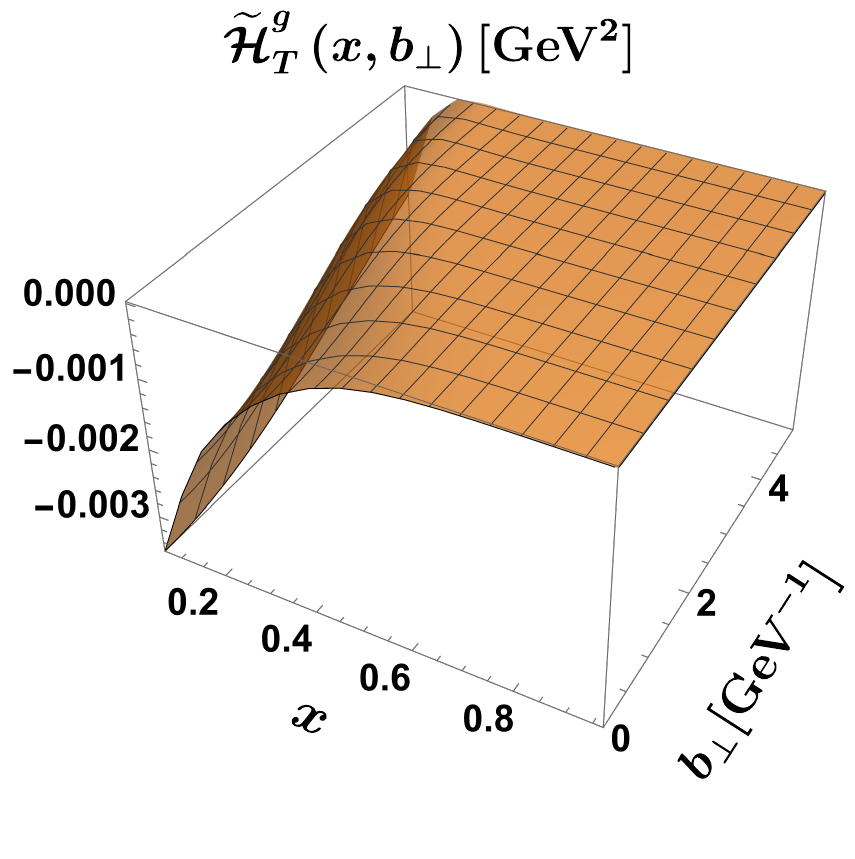}
		\caption{The chiral-odd gluon GPDs in the transverse impact-parameter space, left: $\mathcal{H}_T^g(x,b_{\perp})$, middle: $\mathcal{E}_T^g(x,b_{\perp})$, and right: $\widetilde{\mathcal{H}}_T^g(x,b_{\perp})$ as functions of $x$ and the transverse impact-parameter $b_{\perp}$. The results are evaluated at $\mathcal{N}_{\mathrm{max}}=9$, $\mathcal{K}=16.5$.}
		\label{gluon GPD in impact parameter space}
	\end{center}
\end{figure*}
\section{Numerical results\label{Sec4}}
For our calculations, we set the truncation limits for transverse and longitudinal directions to $\mathcal{N}_{\rm max}=9$ and $\mathcal{K} =16.5$, respectively. The chosen parameters include a harmonic oscillator scale of $b=0.70~\mathrm{GeV}$ and a UV cutoff for instantaneous interactions at $b_{\mathrm{inst}}=3.00~\mathrm{GeV}$. The Hamiltonian model parameters are specified as $\{m_u, m_d, m_g, \kappa, m_f, g_c\} = \{0.31, 0.25, 0.50, 0.54, 1.80, 2.40\}$ (measured in GeV, except for $g_c$). These values were determined to best match the proton's mass, its electromagnetic characteristics, and flavor form factors~\cite{Xu:2023nqv}.

Using these settings, we solve the eigenvalue equation to find the boost-invariant LFWFs, represented by $\Psi^{M_J}_{N,\{\lambda_i\}}(\{x_i,p_i^{\perp}\})$ in Eq.~\eqref{wfs}, at the initial scale of $\mu_0^2=0.23 \sim 0.25~\mathrm{GeV}^2$. We then apply these LFWFs to compute the chiral-odd gluon GPDs in both momentum and impact parameter spaces.

\subsection{The gluon GPDs in momentum space\label{Sec4.1}}
In this section, we detail our findings for the chiral-odd GPDs with zero skewness in momentum space. From this point on, we omit the skewness parameter and express the GPDs as functions of the longitudinal momentum fraction \(x\) and the square of transverse momentum transfer \(-t\). Figure~\ref{gluon GPD in momentum space} illustrates the three nonzero chiral-odd gluon GPDs, \(H_T^g(x,t)\), \(E_T^g(x,t)\), and \(\widetilde{H}_T^g(x,t)\), plotted against \(x\) and \(-t\). Among these, \(H_T^g(x,t)\) exhibits a positive trend, while \(E_T^g(x,t)\) and \(\widetilde{H}_T^g(x,t)\) demonstrate negative values. Our results differ significantly from those obtained using the quark target model (QTM)~\cite{Meissner:2007rx} and light-cone spectator model (LCSM)~\cite{Chakrabarti:2024hwx}, as we observe that the signs of \(H_T^g(x,t)\) and \(E_T^g(x,t)\) in our BLFQ analysis are reversed. Additionally, in our study, \(\widetilde{H}_T^g(x,t)\) is distinctly non-zero, differing from QTM and LCSM results.  
We observe a decrease in the magnitude of the GPD \(H_T^g(x,t)\), along with its peak shifting towards larger $x$ values as the momentum transfer $-t$ increases. This trend mirrors observations made for chiral-even gluon GPDs in our BLFQ approach~\cite{Lin:2023ezw} as well as quark GPDs in the proton in various theoretical investigations~\cite{Ji:1997gm,Scopetta:2002xq,Petrov:1998kf,Penttinen:1999th,Boffi:2002yy,Boffi:2003yj,Vega:2010ns,Chakrabarti:2013gra,Mondal:2015uha,Chakrabarti:2015ama,Mondal:2017wbf,deTeramond:2018ecg,Xu:2021wwj,Liu:2022fvl,Kaur:2023lun}. At zero momentum transfer,  the peak of \(H_T^g(x,t)\) appears at \(x \approx 0.12\). Both \(E_T^g(x,t)\) and \(\widetilde{H}_T^g(x,t)\) display a sharp negative peak at \(-t =0\), with the distribution largely centered around this point. At zero momentum transfer, the peaks of \(E_T^g(x,t)\) and \(\widetilde{H}_T^g(x,t)\) appear at \(x \approx 0.06\) (see insets to the center and right plots in Fig. 1), with the magnitude of \(E_T^g(x,t)\) significantly surpassing those of the other GPDs.

In Fig.~\ref{transverse asymmetry}, we present the transversity asymmetry: $\delta^i \mathcal{J}^i(x)=x [H_T(x,0)+E_T(x,0)+2\widetilde{H}_T(x,0)]$, as outlined in Eq.~\eqref{gluon ta}, plotted against the longitudinal momentum fraction \(x\). We observe that this distribution for both \(u\) and \(d\) quarks is positive, whereas the gluon's distribution is primarily negative for $x<0.165$. 
 Notably, the \(u\) quark's contribution peaks more significantly than that of the \(d\) quark, reaching a peak value of $\sim1.32$ at \(x \sim 0.32\). The peak for the \(d\) quark occurs at a slightly lower \(x\) value of $\sim0.25$, with a peak value of $\sim0.49$. The gluon transversity asymmetry distribution hits a negative peak at \(\sim-0.36\) when \(x\sim0.06\). Interestingly, the gluon distribution changes sign around \(x \sim 0.16\), turning positive thereafter.

\begin{table}[ht]
	\centering
	\caption{Transversity asymmetry for the $u$-quark, $d$-quark, and gluon in the proton. We compare our results for quarks with the predictions from HOM~\cite{Pasquini:2005dk}, HCQM~\cite{Pasquini:2005dk} and $\chi$QSM~\cite{Wakamatsu:2008ki}.}
	\vspace{0.5em}
	\label{table_compare}
	\begin{tabular}{c c c c c}
	\toprule
	\makecell{Transversity\\asymmetry} & \makecell{This work\\BLFQ} & HOM & HCQM & $\chi$QSM \\
	\midrule
		$\braket{\delta^i J^i_u}$ & 0.44  & 0.68 & 0.39 & 0.49 \\
		$\braket{\delta^i J^i_d}$ & 0.14  & 0.28 & 0.10 & 0.22 \\
		$\braket{\delta^i J^i_g}$ & -0.01  &  &  &  \\
	\bottomrule
	\end{tabular}
\end{table}

In Table~\ref{table_compare}, we present the integrated values of the transversity asymmetry distribution over \(x\), alongside a comparison with predictions from other models including the harmonic oscillator model (HOM)~\cite{Pasquini:2005dk}, the hypercentral constituent quark model (HCQM)~\cite{Pasquini:2005dk} , and the chiral quark soliton model ($\chi$QSM)~\cite{Wakamatsu:2008ki}. 
Our findings for both the \(u\) and  \(d\) quarks' contributions fall between those predicted by the HCQM  and $\chi$QSM.
The observed discrepancies may stem from the impact of the initial model scale on the results.
 
Figure~\ref{Fig:moments} depicts our findings for the tensor form factors, which represent the first and second moments of the chiral-odd GPDs, as outlined in Eq.~\eqref{mellin}. 
Specifically, the upper panel of Fig.~\ref{Fig:moments} illustrates the first and second moments of \(H_T^g(x,t)\) across varying values of \(-t\). It is important to note that, in contrast to quarks, where the forward limit of \(H_T^g(x,t)\) aligns with the transversity distribution, a gluon transversity for a spin-$1/2$ system is nonexistent due to helicity conservation constraints \cite{Kumano:2020gfk}. The first and second moments of \(H_T^g(x,t)\) at zero momentum transfer are \(A_{T10}^{g}(0) = 0.067\) and \(A_{T20}^{g}(0) = 0.016\), respectively. In the lower panel Fig.~\ref{Fig:moments}, we present the first two moments of the sum \(E_T^g(x,t) + 2\widetilde{H}_T^g(x,t)\) as a function of \(-t\). Our analysis reveals \(B_{Tn0}^{g}\) as negative in regions close to zero momentum transfer, with values \(B_{T10}^{g}(0) = -0.360\) and \(B_{T20}^{g}(0) = -0.017\). The behavior of \(B_{Tn0}^{g}\) demonstrates both negative and positive regions in $-t$.
 
\begin{figure*}[htp]
	\begin{center}
		\includegraphics[width=0.325\linewidth]{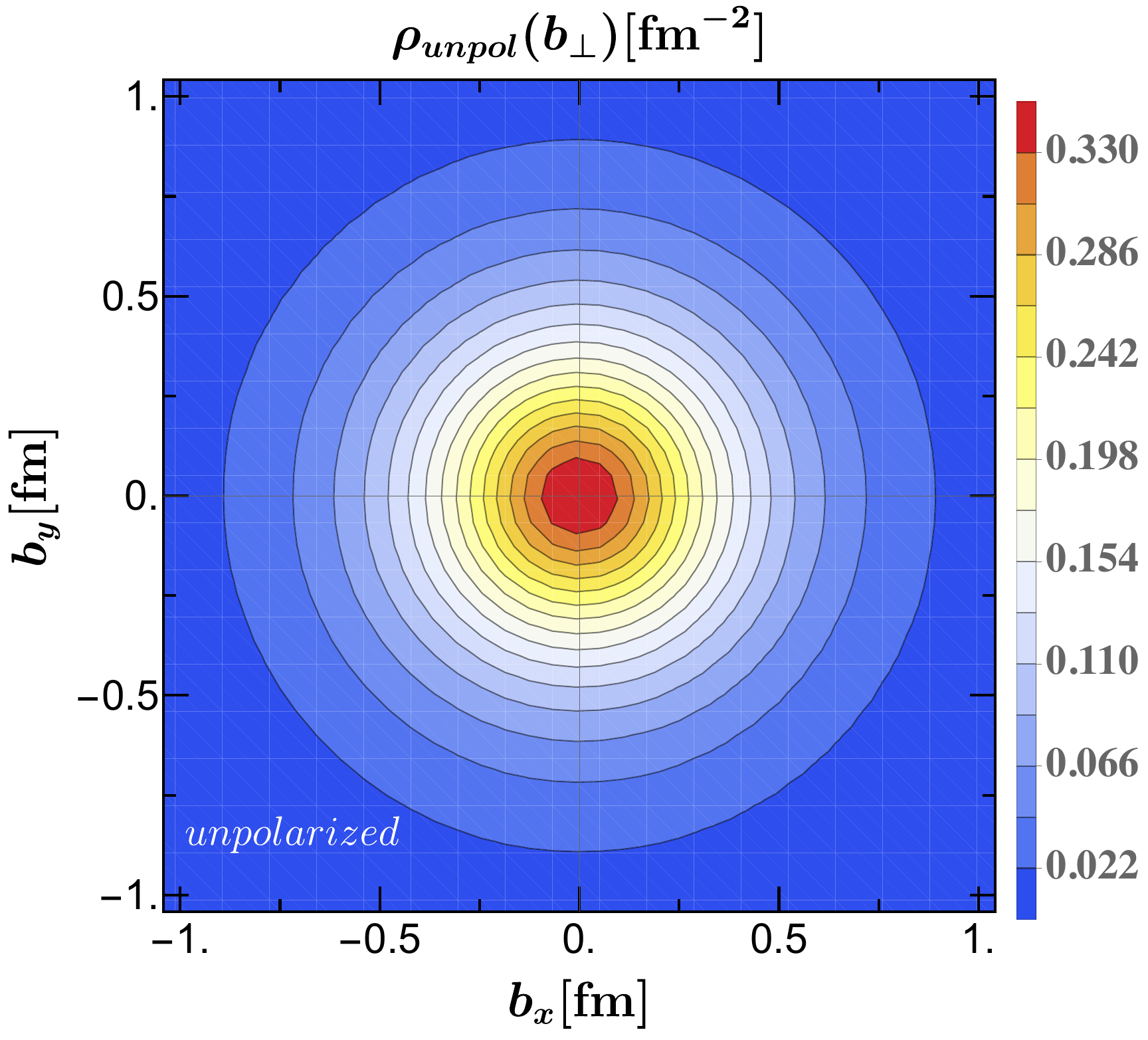}
		\includegraphics[width=0.337\linewidth]{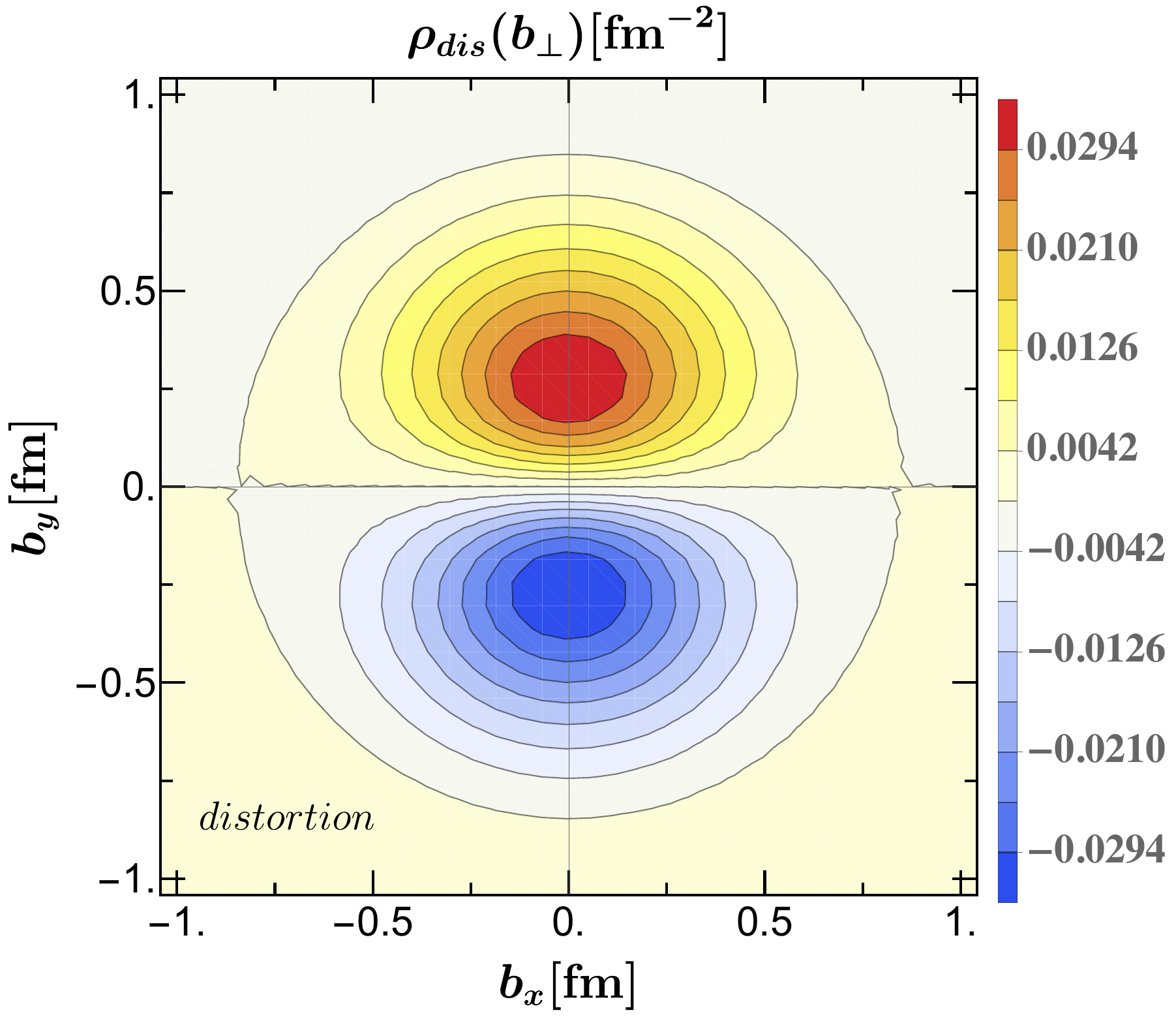}
		\includegraphics[width=0.325\linewidth]{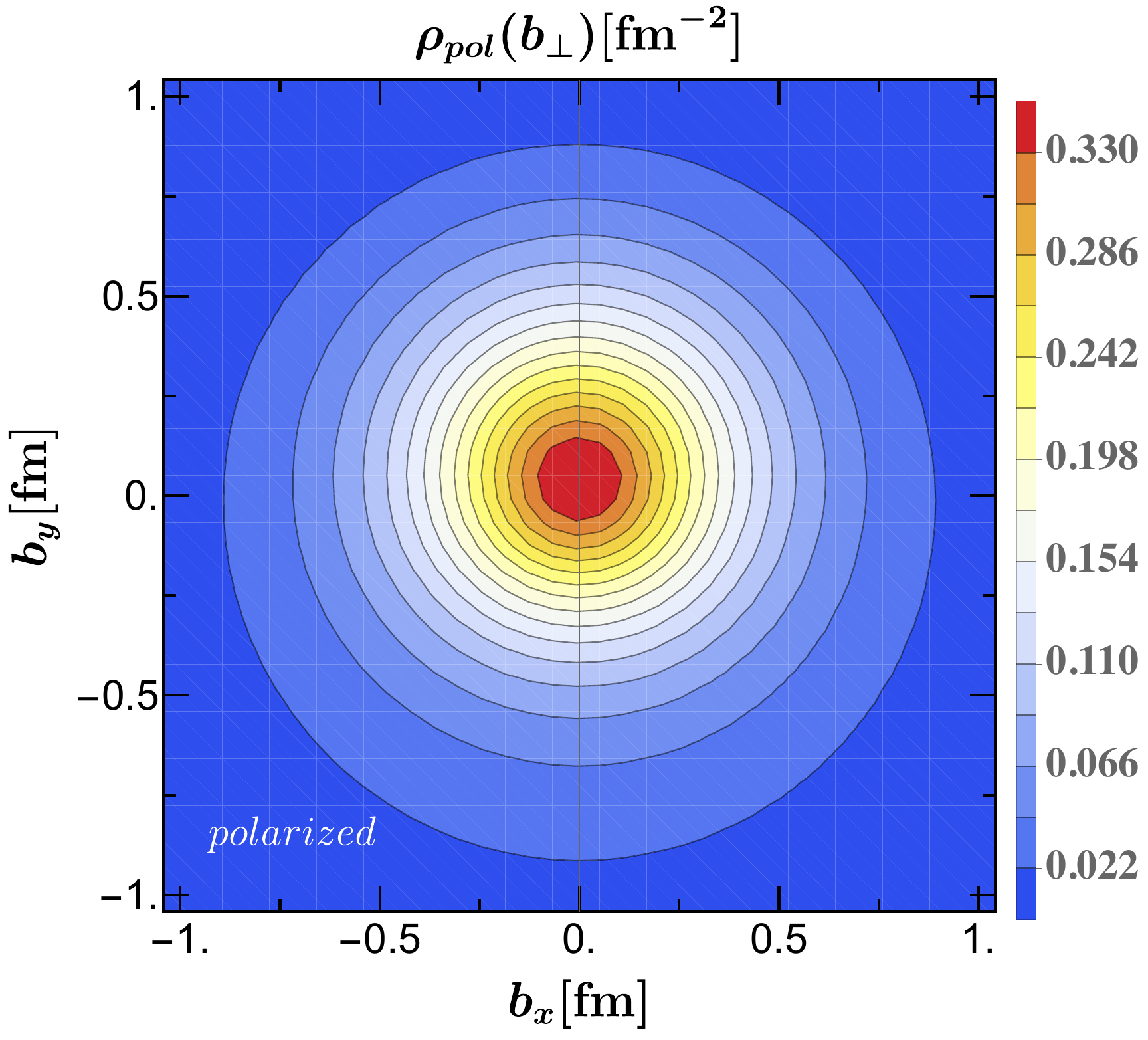}
		\caption{The $x$-integrated gluon spin densities: the unpolarized spin density $\rho_{\mathrm{unpol}}$ (left panel), the spin distortion $\rho_{\mathrm{dis}}$ (middle panel), and the polarized spin density $\rho_{\mathrm{pol}}$ (right panel). The results are evaluated at $\mathcal{N}_{\mathrm{max}}=9$, $\mathcal{K}=16.5$.}
		\label{gluon transverse polarization}
	\end{center}
\end{figure*}
\subsection{The gluon GPDs in the impact parameter space\label{Sec4.2}}
The GPDs in the impact parameter space offer insights into the distribution of partons, characterized by a specific longitudinal momentum fraction $x$ and a transverse position $\vec{b}_{\perp}$ relative to the nucleon's center of momentum. For zero skewness, these distributions in impact parameter space can be interpreted as spin density distributions~\cite{Burkardt:2002hr}. In Fig.~\ref{gluon GPD in impact parameter space}, we illustrate our results for the ipdpdfs, showing them as functions of $x$ and ${b}_{\perp}$. Here, the GPD $\mathcal{H}^g_T(x,b_{\perp})$ is positive, $\widetilde{\mathcal{H}}^g_T(x,b_{\perp})$ is negative, and $\mathcal{E}^g_T(x,b_{\perp})$ is negative in the small $x$ region. We observe peaks in all distributions positioned at the center of the proton ($b_\perp = 0$) when the gluon carries small longitudinal momentum. As $x$ approaches unity, all GPDs converge to zero. We also note that irrespective of their signs, all the GPDs exhibit a reduction in the width of the gluon distributions across the transverse impact-parameter plane as $x$ increases. This suggests that as the gluon acquires a larger longitudinal momentum, they tend to localize nearer to the center of the transverse position. Conversely, as we move away from the center, the distribution peaks shift towards lower $x$ values. Eventually, with increasing transverse distance, the distribution diminishes. Our impact-parameter dependent gluon GPDs exhibit similarities to those analyzed for quarks in several previous studies~\cite{Burkardt:2002hr,Chakrabarti:2013gra,Vega:2010ns,Mondal:2015uha,Chakrabarti:2015ama,Mondal:2017wbf,Maji:2017ill,Kaur:2023lun,Liu:2022fvl}, indicating a trend towards parton and model-independent characteristics.

We further explore the gluon ipdpdfs by considering their spin density, which can be expressed as follows ~\cite{Burkardt:2005hp,Diehl:2005jf,Pasquini:2007xz,Mondal:2018wbl,Maji:2017ill},
\begin{equation}
	\begin{aligned}
		\label{eq_sd}
		\rho \left(x, {b}_{\perp}\right) &=\frac{1}{2}\left[\mathcal{H}\left(x, {b}_\perp\right) - \frac{\varepsilon^{i j}s^i}{M} \frac{\partial}{\partial b_j}\mathcal{B}_T \left(x, {b}_\perp\right)  \right],
	\end{aligned}
\end{equation}
where $\mathcal{B}_T \left(x, {b}_\perp\right) = 2 \tilde{\mathcal{H}}_T\left(x, {b}_{\perp}\right)+\mathcal{E}_T\left(x, {b}_{\perp}\right)$ and $\epsilon^{ij}$ is the two-dimensional antisymmetric tensor, with $\epsilon^{12}=1$ and $\epsilon^{21}=-1$. $\mathcal{H}\left(x, {b}_\perp\right)$ represents the unpolarized chiral even gluon GPD, which was previously calculated in Ref.~\cite{Lin:2023ezw}. For the unpolarized spin density ($\rho_{\mathrm{unpol}}$), the polarization vector $s^i = 0$ and the polarized spin density ($\rho_{\mathrm{pol}}$) is obtained when $s^1 =1, s^2=0$. The second term in Eq.~\eqref{eq_sd} provides the distortion ($\rho_{\mathrm{dis}}$) in the density  caused by the polarization. In Fig.~\ref{gluon transverse polarization}, we display the contour plot of the $x$-integrated spin densities for $\rho_{\mathrm{unpol}}$, $\rho_{\mathrm{dis}}$, and $\rho_{\mathrm{pol}}$ with respect to the two transverse directions in the impact parameter space $b_x$ and $b_y$. The term $\mathcal{B}_T$ behaves similarly to the chiral even GPD $E$, thus causing the transverse distortion when the gluon polarization is present inside the unpolarized proton. This distortion in the IPS can be observed in Fig.~\ref{gluon transverse polarization} (third panel) where the distribution is shifted upwards compared to the unpolarized case (first panel), where the distribution is rotationally symmetric in the IPS. The center panel in Fig.~\ref{gluon transverse polarization} illustrates the distortion caused by $ \frac{\partial}{\partial b_j}\mathcal{B}_T$, which has a dipolar nature.

We quantify the average transverse shift of the peak position of the spin density along the $b_y$-direction, regarding a gluon's transverse spin aligned in the $x$-direction, as defined by~\cite{QCDSF:2007ifr},
\begin{align}
    \langle b^{y}_\perp\rangle^n=\frac{\int {\rm d}x\,{\rm d}^2 b_\perp\,  x^{n-1} b^y_\perp\rho_{\rm pol} \left(x, {b}_{\perp}\right)}{\int {\rm d}x\,{\rm d}^2 b_\perp\, x^{n-1}\rho_{\rm unpol} \left(x, {b}_{\perp}\right)}.
\end{align}
Our BLFQ results give: $\langle b^{y}_\perp\rangle^1=0.097$ fm and $\langle b^{y}_\perp\rangle^2=0.015$ fm.


\section{Summary and outlook\label{Sec5}}
Basis Light-front Quantization (BLFQ) is a recently developed nonperturbative framework for solving quantum field theory. In this work, we have calculated the chiral-odd gluon GPDs of the proton
from its light-front wave functions within the framework of BLFQ. 
These wave functions have been obtained from the eigen solutions of the light-front QCD Hamiltonian in the light-cone gauge for the proton expressed within  $|qqq\rangle$ and $|qqqg\rangle$ Fock spaces, together with a three-dimensional confinement in the leading Fock
sector.

We have conducted an analysis of chiral-odd gluon GPDs at zero skewness, examining them in both momentum space and impact parameter space. Our findings reveal that $H_T^g$ displays a positive distribution, whereas $E_T^g$ and $\widetilde{H}_T^g$ exhibit negative distributions, contrasting significantly with results from the quark target model (QTM)~\cite{Meissner:2007rx} and light-cone gluon-spectator model (LCSM)~\cite{Chakrabarti:2024hwx}. Notably, we have observed a reversal in the signs of $H_T^g$ and $E_T^g$ compared to the QTM and LCSM results in our BLFQ analysis. Furthermore, our study reveals a non-zero $\widetilde{H}_T^g$, setting it apart from those phenomenological models' predictions.

We have computed the gluon generalized form factors for the proton, specifically the tensor form factors derived from the first two moments of gluon tensor GPDs. These second moments of chiral-odd GPDs offer detailed insights into gravitational form factors and are associated with the angular momentum carried by the gluon with transverse spin  in an unpolarized proton. Our analysis reveals that the form factors $A_{Tn0}^g(t)$, derived from the moments of the GPD $H_T^g(x,t)$, exhibit a positive trend and decrease monotonically as $-t$ increases. Conversely, the form factors $B_{Tn0}^{g}$, derived from the combination $E_T^g(x,t) + 2\widetilde{H}_T^g(x,t)$, are negative in regions near zero momentum transfer but become positive as $-t$ increases, eventually approaching zero at large $-t$. At our model scale, we obtain $A_{T20}^{g}(0) = 0.016$ and $B_{T20}^{g}(0) = -0.017$. Following the proton transverse spin sum rule, we have illustrated the $x$-dependence of the angular momentum carried by polarized gluons and determined the relative contributions of quarks and gluons to the transversity asymmetry. We have observed positive quark transversity asymmetries, consistent with other phenomenological models, while our approach indicates a negative and relatively smaller gluon transversity asymmetry.

We have subsequently computed the probability densities of both unpolarized and polarized gluons within the proton. We have found that the spatial distribution of unpolarized gluons is axially symmetric. However, when the gluon is  polarized, the distribution becomes strongly distorted, indicating a nontrivial pattern of gluon polarization within the proton. To quantify the shift in peak positions of these densities, we have calculated the average transverse shift and obtained $\langle b^{y}_\perp\rangle^1=0.097$ fm and $\langle b^{y}_\perp\rangle^2=0.015$ fm.

Expanding our analysis to incorporate non-zero skewness is essential, particularly as experiments explore GPDs at non-zero skewness. Since our BLFQ approach is numerical, it requires the application of specific interpolation techniques to ensure accurate overlaps of wave functions across various longitudinal momentum fractions. This interpolation-based methodology has proven successful in examining quark skewed GPDs~\cite{Liu:2024umn}. Future research will focus on gluon GPDs at non-zero skewness and will extend to higher Fock states, with the goal of investigating GPDs for sea quarks within the proton.


\section*{Acknowledgements\label{Sec6}}
We thank Zhimin Zhu, Ziqi Zhang and Yiping Liu for helpful discussions in the Institute of Modern Physics, University of Chinese Academy of Science. S. N. is supported by the Senior Scientist Program funded by Gansu Province, Grant No. 23JDKA0005. CM thanks the Chinese Academy of Sciences President’s International Fellowship Initiative for the support via Grants No. 2021PM0023. C. M. is also supported by new faculty start up funding by the Institute of Modern Physics, Chinese Academy of Sciences, Grant No. E129952YR0. X. Z. is supported by new faculty startup funding by the Institute of Modern Physics, Chinese Academy of Sciences, by Key Research Program of Frontier Sciences, Chinese Academy of Sciences, Grant No. ZDBS-LY-7020, by the Natural Science Foundation of Gansu Province, China, Grant No. 20JR10RA067, by the Foundation for Key Talents of Gansu Province, by the Central Funds Guiding the Local Science and Technology Development of Gansu Province, Grant No. 22ZY1QA006, by Gansu International Collaboration and Talents Recruitment Base of Particle Physics (2023-2027), by International Partnership Program of the Chinese Academy of Sciences, Grant No. 016GJHZ2022103FN and by the Strategic Priority Research Program of the Chinese Academy of Sciences, Grant No. XDB34000000. J. P. V. is supported by the Department of Energy under Grant No. DE-SC0023692. A major portion of the computational resources were also provided by Sugon Advanced Computing Center.

\biboptions{sort&compress}
\bibliographystyle{elsarticle-num}
\bibliography{ProtonGPDletter.bib}
\end{document}